\documentclass[twocolumn, apl, amsmath, amssymb, showpacs]{revtex4-2}
\usepackage{epsf}      
\usepackage{graphicx}
\usepackage{color}
\usepackage{soul}
\usepackage{gensymb}
\usepackage{sidecap}
\usepackage{amsmath}
\usepackage{mathtools}
\usepackage[colorlinks=true, linkcolor=blue, citecolor=blue, urlcolor=blue]{hyperref}
\begin{document}

\title{Non-trivial critical behavior at the magnetic transitions: A case study of Sm$_7$Pd$_3$}

\author{Ajay Kumar}

\email{ajay1@ameslab.gov}

\author{Anis Biswas}

\author{Y. Mudryk}

\affiliation{Ames National Laboratory, U.S. Department of Energy, Iowa State University, Ames, Iowa 50011, USA} 

\date{\today}

\begin{abstract}

We present a comprehensive analysis of the critical behavior of Sm$_7$Pd$_3$ in the vicinity of its second-order magnetoelastic transition at $T_ {\rm c} = 173$ K. The critical exponents (CEs) $\beta$ and $\gamma$, determined using both the standard convergence procedure and the average normalized slope (ANS) method, diverge at $T_{\rm c}$: a characteristic typically associated with first-order transitions. Notably, none of the  established universality classes satisfactorily describe the critical behavior of Sm$_7$Pd$_3$, and we discuss the possible origins of this deviation in the context of the strong spin-lattice coupling intrinsic to the sample. We emphasize the importance of accurately selecting the critical temperature and magnetic field ranges to ensure robust critical behavior analysis and propose a quantitative approach to assess the reliability of the extracted CEs. Additionally, we demonstrate that in the ANS method, the critical exponents $\beta$ and $\gamma$ should be calculated separately using data for $T \leqslant T_{\rm c}$ and $T \geqslant T_{\rm c}$, respectively. Our findings underscore the need for a revised theoretical framework to accurately describe second-order magnetoelastic transitions.

\end{abstract}

\maketitle

\section{\noindent ~Introduction}

Solid-state phase transitions exemplify cooperative behavior of matter near critical points, one of the focal areas of condensed matter physics and physical chemistry ~\cite{Narayan_NM_19, Aguiar_PRL_09, Chen_PRL_19}. Detailed investigation of magnetic ordering transitions can often provide deep insights into the nature of magnetic interactions in the ordered state ~\cite{Quintana_PRL_23, Sarkar_PRL_09, Butch_PRL_09, Biswas_PRM_24}. In particular, the transition from the paramagnetic (PM) to the ferromagnetic (FM) state has been extensively investigated across a wide range of materials for this purpose. When the PM--FM transition is second order, the behavior of the magnetic moments in the asymptotic region--near the transition temperature ($T_{\rm c}$)--can be described using equations of state characterized by a set of critical exponents (CEs)~\cite{Fisher_RMP_85, Fisher_1972, Liu_PRB_18}. The critical exponent $\beta$ determines the nature of magnetic interactions in the ordered FM state, whereas $\gamma$ governs the decay of the ordered moment in the PM state, within the asymptotic region. Therefore, the values of these  exponents define the universality class of the magnetic ground state in a material, providing a general description of its magnetic behavior \cite{Liu_PRB_17, Campostrini_PRB_02, Pelissetto_PR_02}. This is particularly valuable for systems where conventional techniques such as neutron diffraction are impractical due to  intrinsically low neutron scattering cross-sections of constituent elements. In contrast, when the PM--FM transition is first order and involves a structural phase change, the standard scaling relations fail, and the critical exponents diverge at $T_{\rm c}$~\cite{Fisher_RMP_67, Kumar_JAP_24}. However, there exists an intriguing class of materials that exhibit either first or second-order PM--FM transition with no change in crystal symmetry and structure type but with notable changes in lattice parameters, which we refer to as “magnetoelastic transition”. Such behavior is frequently observed in rare-earth intermetallic compounds~\cite{Tartaglia_PRB_19, Paudyal_PRB_08, Kumar_PRB_24}  and offers an opportunity to explore the interplay between lattice and magnetic degrees of freedom in the absence of conventional magnetostructural transformations.

A wide range of materials exhibit anomalous critical behavior, wherein the extracted set of critical exponents does not conform to any established universality class of FM systems ~\cite{Chauhan_PRL_22, Meng_PRB_23, Pandey_PRB_24, Yadav_SR_21}. This deviation may arise either due to underlying structural modifications at the transition, which are not fully captured by conventional scaling theories, or due to the presence of competing or non-FM interactions in the ordered state.  A systematic understanding of the origins of such deviations remains elusive, particularly in systems where magnetism couples strongly with the lattice. Recently, following the theoretical prediction~\cite{Leonard_PRL_15}, Chauhan \textit{et al.} reported a non-trivial critical behavior in the skyrmion-host Cu$_2$OSeO$_3$, where a transition in the universality class from the 3D Ising to the 3D Heisenberg model has been observed above and below  $T_{\rm c}$, respectively~\cite{Chauhan_PRL_22}. A similar transition in CEs has also been observed in Cr$_{1/3}$TaS$_2$~\cite{Meng_PRB_23} and Cr$_{1+x}$Te$_2$ \cite {Pandey_PRB_24}. However, such a transition from one universality class in the FM state to another in the PM state complicates the physical interpretation of the underlying magnetic interactions, necessitating a more detailed investigation from both fundamental and methodological perspectives.

Despite many studies, a systematic understanding of the origins of anomalous critical behavior across second order phase transition in various magnetic materials remains poorly understood. In this context, Sm$_7$Pd$_3$ serves as a prototypical system for studying critical behavior at a second-order magnetoelastic transition, which hosts both complex magnetic interactions \cite{Kumar_PRB_25} and significant lattice change at $T_{\rm c}$ \cite{Biswas_AM_24}. Sm$_7$Pd$_3$ undergoes a PM--FM transition at $T_{\rm c} \approx 170$~K and crystallizes in the Th$_7$Fe$_3$-type non-centrosymmetric hexagonal structure (space group $P6_3mc$) at room temperature, which is retained down to 5~K~\cite{Kadomatsu_JMMM_98, Biswas_AM_24}; however, an abrupt change in the lattice parameters (particularly the $c/a$ ratio) has been observed at the magnetic transition, indicative of strong magnetoelastic coupling~\cite{Biswas_AM_24}. The presence of a $\lambda$-shaped anomaly in the specific heat at $T_{\rm c}$ confirms the second-order nature of this transition~\cite{Kumar_PRB_25}. Despite the observation of giant low-temperature coercivity, suggesting the presence of dominant FM interactions, density functional theory (DFT) calculations indicate an antiferromagnetic (AFM) coupling between Sm atoms located on different atomic planes along the $c$-axis~\cite{Biswas_AM_24}. Our recent study elucidates the formation of a glassy state due to the coexistence of these competing FM and AFM interactions~\cite{Kumar_PRB_25}. Given the absence of neutron powder diffraction (NPD) data due to the low coherent neutron scattering cross-section of Sm, the investigation of the critical behavior of Sm$_7$Pd$_3$ can serve as a useful tool to unravel the true nature of magnetic interactions in this system.

In this work, we present a comprehensive critical behavior analysis of Sm$_7$Pd$_3$ using two different approaches. Our analysis demonstrate that none of the established universality classes provide a satisfactory convergence of the critical exponents for this compound. We discuss potential origins of the observed non-trivial critical behavior in Sm$_7$Pd$_3$, in the broader context of similar anomalies reported in other magnetic materials, and propose modifications to the existing frameworks for critical behavior analysis in such systems.

\begin{figure*}
\centering
\includegraphics[width=1\textwidth]{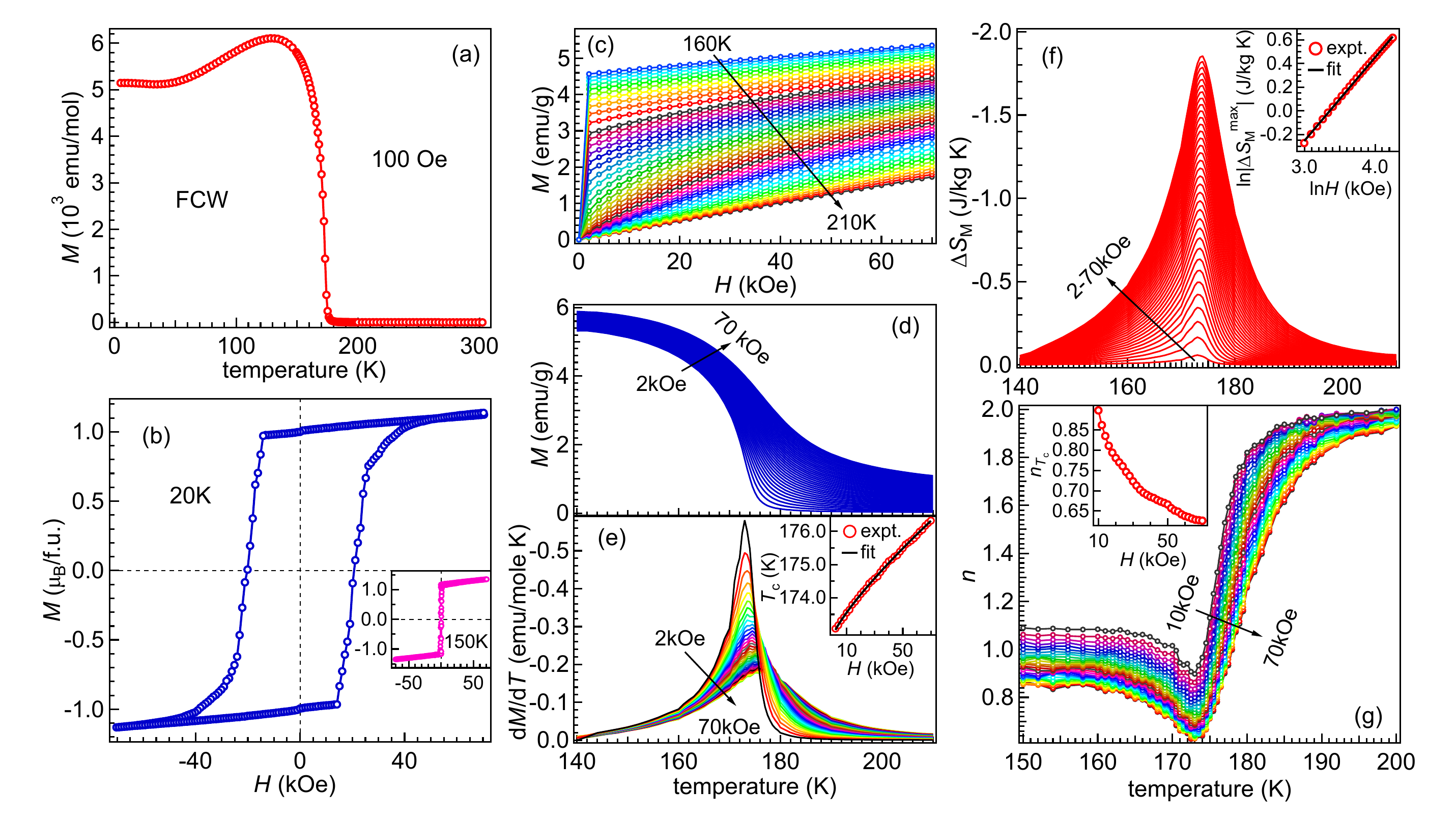}
\caption {(a) Temperature-dependent magnetization of Sm$_7$Pd$_3$ measured in the field-cooled warming (FCW) mode at 100~Oe. (b) Zero field cooled (ZFC) isothermal magnetization ($M$--$H$) curve at 20~K; inset shows the ZFC $M$--$H$ at 150~K. (c) Virgin magnetization isotherms recorded at various temperatures across the magnetic transition. (d) Temperature-dependent magnetization at different magnetic fields derived from the virgin isotherms. (e) Temperature derivative of magnetization at different fields; inset shows the shift in $T_{\rm c}$ as a function of external field, where the solid black curve represents the power-law fit. (f) Temperature-dependent magnetic entropy change ($\Delta S_{\rm M}$) at various magnetic fields; inset shows the maximum $\Delta S_{\rm M}$ as a function of field on a log--log scale. (g) Local field exponent $n$ as a function of temperature at different fields; inset shows the field dependence of $n$ at $T_{\rm c}$.} 
\label{Fig1_MCE}
\end{figure*}

\section{\noindent ~experimental}

The polycrystalline Sm$_7$Pd$_3$ sample was prepared by arc melting stoichiometric amounts of high-purity Sm (99.95 wt.\%) from the Materials Preparation Center of Ames National Laboratory and Pd (99.99+ wt.\%). The resulting sample was confirmed to be single-phase, crystallizing in the Th$_7$Fe$_3$-type hexagonal structure. We also note that all other reported binary phases in the Sm-Pd system are either diamagnetic or Pauli paramagnetic \cite{Jordan_JLCM_75}; therefore, their minor presence below the detection limit of XRD is unlikely to influence the observed magnetic behavior. Further details on the sample preparation, as well as its structural and magnetic characterization, can be found elsewhere~\cite{Biswas_AM_24, Kumar_PRB_25}. Temperature-dependent magnetization ($M$--$T$) measurements were carried out in the field-cooled warming mode using a superconducting quantum interference device (SQUID) magnetometer (MPMS XL-7, Quantum Design, USA). The field-dependent magnetization ($M$--$H$), and virgin isotherms at various temperatures were recorded  across the magnetic transition ($\Delta T$ = 0.2~K) using the high-precision reciprocating sample option (RSO) of the same SQUID system in the ``No Overshoot'' field ramping mode. To eliminate remanent magnetization, the sample was heated to room temperature after  recording the $M$-$H$ curves at each temperature. However, due to low coercivity ($<$100~Oe) of the sample in the critical regime (near $T_{\rm c}$), the magnetic field was reduced from 70~kOe to 500~Oe linearly, and then from 500~Oe to 0~Oe in the oscillatory mode while recording the successive virgin isotherms.  All data plotting and analysis were performed using the \textit{Igor Pro 9.02} software.

\begin{figure*}
\centering
\includegraphics[width=5.5in]{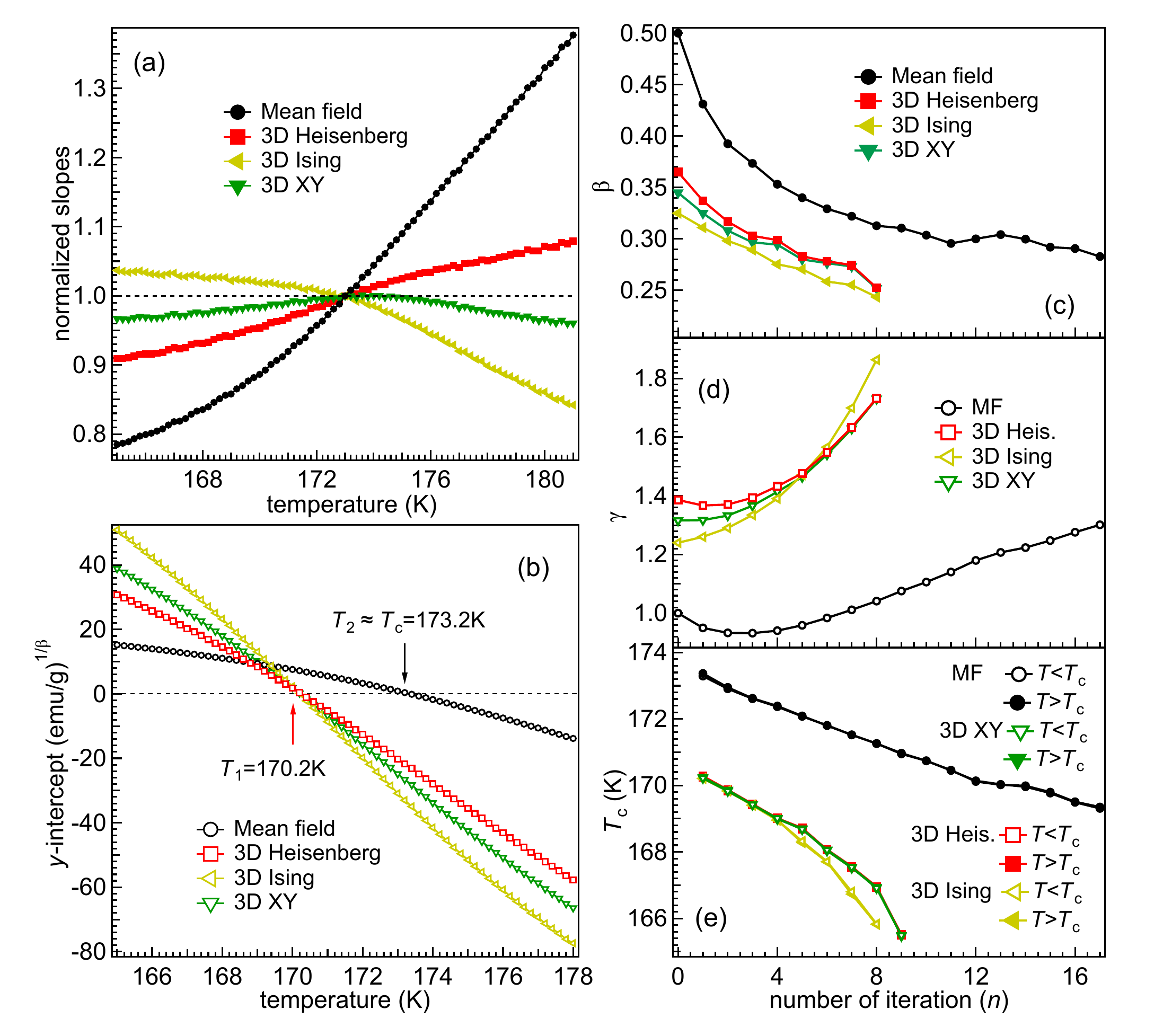}
\caption {(a, b) Normalized slopes (NS) and $y$-intercepts of the modified Arrott plots (MAPs) for different universality classes.  (c--e) Variation of the critical exponents $\beta$, $\gamma$, and the transition temperature with the number of iterations, starting from different universality classes.} 
\label{Fig2_SCP}
\end{figure*}

\section{\noindent ~Results and discussion}

The temperature-dependent magnetization of Sm$_7$Pd$_3$, recorded in the field-cooled warming mode at 100~Oe, is presented in Fig.~\ref{Fig1_MCE}(a). A sharp PM to FM-like transition is observed at $\approx$173~K, consistent with previous reports~\cite{Kadomatsu_JMMM_98, Biswas_AM_24, Kumar_PRB_25}. Moreover, a decrease in magnetization, which persists even in the high magnetic fields, is observed at low temperatures (below $\sim$130~K) \cite{Biswas_AM_24, Kumar_PRB_25}, suggesting the presence of finite AFM interactions in the sample. Figure~\ref{Fig1_MCE}(b) shows the field-dependent magnetization at 20~K, which exhibits large coercivity ($H_{\rm c}\sim$20~kOe), indicating dominant FM coupling in the sample. The inset shows the $M$-$H$ curve at 150~K, highlighting the soft FM behavior of the sample ($H_{\rm c}<$ 1~kOe) in this temperature range. To understand the nature of magnetic interactions in the magnetically ordered state, virgin magnetization isotherms were recorded at various temperatures across the magnetic transition, as shown in Fig.~\ref{Fig1_MCE}(c). The $M$--$T$ curves derived from these magnetization isotherms are presented in Fig.~\ref{Fig1_MCE}(d) at different fields. The transition becomes broader and shifts to higher temperatures with increasing magnetic field, as expected for a PM--FM transition. This behavior is more clearly observed from the temperature derivative curves shown in Fig.~\ref{Fig1_MCE}(e). The d$M$/d$T$ curves exhibit a sharp peak, which was fitted using a Gaussian line shape to precisely determine the transition temperature, plotted in the inset of Fig.~\ref{Fig1_MCE}(e).

A linear increase in $T_{\rm c}$ with external magnetic field has been widely observed in materials exhibiting a first-order PM--FM transition~\cite{Kumar_JAP_24, Pecharsky_PRL_97, Guillou_NC_18}. In contrast, for second-order phase transitions (SOPT), both theoretical~\cite{Sznajd_PRB_01} and experimental studies~\cite{Kokorin_JMMM_17, Schwarz_JMMM_11} have shown that the temperature corresponding to the peak in magnetic susceptibility (identified as $T_{\rm c}$) follows a power-law dependence on the applied magnetic field, typically expressed as:
\begin{equation}
T_{\rm c}(H) = T_{\rm c}(0) + A H^\phi,
\label{power}
\end{equation}
where $T_{\rm c}(0)$ is the Curie temperature in zero field, $A$ is a material-specific amplitude factor, and $\phi$ is the power exponent. Within the Landau mean-field (MF) framework for SOPT, $\phi = 2/3$~\cite{Sznajd_PRB_01}, while $\phi = 1$ is typically associated with first-order transitions. In the case of Sm$_7$Pd$_3$, the $T_{\rm c}(H)$ data exhibit a clear non-linear trend [see inset of Fig.~\ref{Fig1_MCE}(e)], which is well described by the power-law relation given in Eq.~(\ref{power}). The best fit, shown as a black solid curve in the inset of Fig.~\ref{Fig1_MCE}(e), yields $T_{\rm c}(0) = 172.84(2)$~K, $A$ = 4.2$\times$10$^{-4}$ K Oe$^{-0.81}$, and $\phi = 0.81(1)$. The value of $\phi$ lies between the first-order and mean-field SOPT limits, which suggests a non-trivial interplay between the magnetic and lattice degrees of freedom, likely arising from strong magnetoelastic coupling in Sm$_7$Pd$_3$~\cite{Biswas_AM_24}.


Further, we calculate the magnetic entropy change ($\Delta S_{\rm M}$) of the sample using the following Maxwell relation~\cite{Phan_JMMM_07, Kumar_PRB_20}:

\begin{eqnarray} 
\Delta S_{\rm M}(T, \Delta H) &=& \int_0^H \left( \frac{\partial M(T,H)} {\partial T} \right)_H {\rm d}H
\end{eqnarray}

The $\Delta S_{\rm M}(T)$ curves are shown in Fig.~\ref{Fig1_MCE}(f) at different magnetic fields, exhibiting a sharp peak centered around 173~K that enhances monotonically with the applied field. For second-order transition materials, the maximum magnetic entropy change ($\Delta S^{\rm max}_{\rm M}$) exhibits a power-law dependence on the magnetic field, i.e., $\Delta S^{\rm max}_{\rm M} \propto \Delta H^n$ \cite{Franco_APL_06}.  Therefore, in the inset of Fig.~\ref{Fig1_MCE}(f), we plot $\ln H$ versus $\ln |\Delta S^{\rm max}_{\rm M}|$, which displays linear behavior. The best fit using a power law for $H = 20$--70~kOe, shown by the solid black line, gives $n = 0.705(4)$, which is close to the theoretical value of 2/3 for the mean-field model~\cite{Franco_APL_06}. Furthermore, the local value of the power exponent is calculated using $n(T, H) = \frac{{\rm d} \ln |\Delta S_{\rm M}|}{{\rm d} \ln H}$, as shown in Fig.~\ref{Fig1_MCE}(g). The value of $n$ is less than 2 for all temperatures and magnetic fields, confirming the second-order nature of this transition~\cite{Law_NP_18}. However, the value of $n$ at $T_{\rm c}$ decreases monotonically with an increase in the magnetic field, as shown in the inset of Fig.~\ref{Fig1_MCE}(g), which prevents us from reliably assigning the universality class and, hence, the nature of magnetic interactions to the system.  Moreover, the critical exponents extracted from the different scaling relations of the $\Delta S_{\rm M}(T, H)$ curves yield unphysical values, despite the clear second-order nature of the magnetic transition in Sm$_7$Pd$_3$ [see Figs. S1(a--c) of Ref.~\cite{SI} and the discussion therein].

\begin{table*}
    \centering
    \caption{The critical exponents extracted using different methods and their comparison with the literature and theoretical values for the different universality classes.}
    \begin{tabular}{p{2.5cm}p{2.5cm}p{2cm}p{2cm}p{2cm}p{2cm}p{2cm}}
        \hline
        Sample/Class & Method & $\beta$ & $\gamma$ & $\delta$ & $T_{\rm c}$ (K)& Reference  \\
        \hline
        & MAP &  0.325(1) & 1.068(5) & 4.286(6) & 171.3(1) & This work\\
        Sm$_7$Pd$_3$ & KF &   0.320(5) & 1.069(4) & 4.340(6) & 171.3(1) & This work\\
        & Critical isotherm &    &  & 4.292(7) &  & This work \\
Gd$_7$Pd$_3$ & KF &   0.34(2) & 1.010(3) & 3.95(10) & 131.8(2) & \cite{Griffiths_PRB_85} \\
Gd$_7$Pd$_3$H$_2$ & KF &   0.37(4) & 1.01(5) & 3.92(8) & 106.3 & \cite{Griffiths_JPFMP_88} \\
        \hline
        Mean field & Theory &  0.5 & 1.0 & 3.0 &  &\cite{Kaul_JMMM_85, Guillou_PRB_80}\\
        3D Heisenberg & Theory  & 0.365 & 1.386 & 4.80 &  &\cite{Kaul_JMMM_85, Guillou_PRB_80}\\
        3D Ising & Theory  & 0.325 & 1.24 & 4.82 &  &\cite{Kaul_JMMM_85, Guillou_PRB_80}\\
        3D XY& Theory  & 0.345 & 1.316 & 4.81 &  &\cite{Kaul_JMMM_85, Guillou_PRB_80}\\   
   
        \hline
    \end{tabular}
    \label{T_critical}
\end{table*}

Therefore,  to gain deeper insight into the nature of FM interactions in Sm$_7$Pd$_3$,  we attempted to determine the critical exponents $\beta$, $\gamma$, and $\delta$, which govern the magnetization of the sample in asymptotic region, using the following equations of state~\cite{Kouvel_PR_64, Kouvel_PRL_68}:

\begin{eqnarray}
\label{c1}
 M_{\rm SP}(0, T) = M_0 (-\epsilon)^{\beta} \quad \text{for } \epsilon < 0, \quad T < T_{\rm c} \\
\label{c2}
 \chi_0^{-1}(0, T) = \left(\frac{h_0}{M_0}\right) \epsilon^{\gamma} \quad \text{for } \epsilon > 0, \quad T > T_{\rm c}  \\
 M(H, T_{\rm c} ) = D H^{1/\delta} \quad \text{for } \epsilon = 0, \quad T = T_{\rm c} 
\label{c3}
\end{eqnarray}

where $\epsilon = (T - T_{\rm c})/T_{\rm c}$, $M_{\rm SP}$ and $\chi_0^{-1}$ represent the spontaneous magnetization and initial inverse susceptibility below and above $T_{\rm c}$, respectively, and $M_0$, $h_0/M_0$, and $D$ are constants. The   field dependence of the magnetization in the asymptotic region of a FM transition is described by the Arrott-Noakes equation given as~\cite{Arrott_PRL_67}

\begin{eqnarray}
\left(\frac{H}{M}\right)^{1/\gamma} = a \epsilon + b M^{1/\beta},
\label{AN}
\end{eqnarray}

where $a$ and $b$ are the temperature-dependent constants. The above equation manifests the linear behavior of the $M^{1/\beta}$ vs. $(H/M)^{1/\gamma}$ curves [modified Arrott plots (MAPs)] in the critical region. The MAPs of Sm$_7$Pd$_3$ constructed using the virgin magnetization isotherms are shown in Figs. S2 (a--d) of \cite{SI} for different universality classes using their corresponding values of the critical exponents, $\beta$ and $\gamma$ (given in Table I). All the plots exhibit a positive slope, which confirms the second-order nature of the FM to PM transition in Sm$_7$Pd$_3$ \cite{Banerjee_PL_64}. In Fig. \ref{Fig2_SCP}(a), we plot the normalized slopes (NS) for different universality classes, where NS is defined as the ratio of the slope of MAP at a given temperature to its slope at $T_{\rm c}$ = 173~K \cite{Kumar_PRB_25}. A linear fit of the MAPs from 20-70~kOe [shown by solid red lines in Figs. S2 (a--d) of \cite{SI}] has been used to calculate the NS. It can be inferred from Fig. \ref{Fig2_SCP}(a) that below $T_{\rm c}$, the 3D Ising and 3D XY models show the least deviation from unity, whereas above $T_{\rm c}$, the 3D XY and 3D Heisenberg models are closest, indicating that none of the universality classes can solely explain the complex magnetic interactions in Sm$_7$Pd$_3$. The $y$-intercept of the MAPs, on the other hand, exhibits a crossover from positive to negative values at $T_1 = 170.2$~K for the 3D Heisenberg, 3D Ising, and 3D XY models (short-range magnetic interactions), and at $T_2 = 173.2$~K for the mean-field model (long-range interactions), as indicated by the red and black arrows, respectively, in Fig.~\ref{Fig2_SCP}(b). A similar behavior, featuring two distinct crossover temperatures of the $y$-intercept, has recently been observed in Ni metal~\cite{Chauhan_arXiv_25}.

As guided by the behavior of NS, we start with the theoretical values of the critical exponents for the 3D XY model ($\beta = 0.345$ and $\gamma = 1.316$) and calculate the $M_{\rm SP}(T)$ and $\chi_0^{-1}(T)$ from the $y$- and $x$-intercepts of the MAP [Fig. S2(d) of \cite{SI}]. Then, by fitting the $M_{\rm SP}(T)$ and $\chi_0^{-1}(T)$ using equations \ref{c1} and \ref{c2}, respectively, a new set of $\beta$ and $\gamma$ is obtained and used to reconstruct the MAP. Again, a linear fit of the high-field region (20-70~kOe) gives a modified set of the critical exponents and hence a new MAP. For a trivial system, the critical exponents $\beta$ and $\gamma$ should converge to the stable values after some iterations \cite{Liu_PRB_18, Zhang_PRB_12, Biswas_PRM_24}. However, in the present case, the value of $\beta$ monotonically decreases, whereas $\gamma$ keeps on increasing with the number of iterations ($n$), as shown in Figs. \ref{Fig2_SCP}(c, d), respectively. The critical temperature ($T_{\rm c}$) also decreases with $n$ for all the universality classes [Fig. \ref{Fig2_SCP}(e)]. This non-trivial diverging behavior of the critical exponents suggests the complex magnetic interactions in Sm$_7$Pd$_3$, which cannot be explained by above mentioned state of equations. Here, it is important to emphasize that both the range of the critical temperatures ($T_{cri}$), magnetic fields ($H_{cri}$) , and their step size are crucial for this analysis.  The $T_{cri}$ is governed by the broadening of the magnetic transition, and the field-dependent shift in the $T_{\rm c}$, whereas the $H_{cri}$ depends on the magnetic hardness of the materials in the critical region. We use a temperature range of $T_{\rm c}$ $\pm$ 5~K (165 -- 175~K for 3D Heisenberg, 3D Ising, and 3D XY, and 168 -- 178~K for MF; $\sim$ $\pm$ 3\% of $T_{\rm c}$) with a temperature step of 0.2~K and a magnetic field range of 20-70~kOe ($\Delta H$ = 2~kOe) to fit the MAPs.

\begin{figure}
\centering
\includegraphics[width=3.5in]{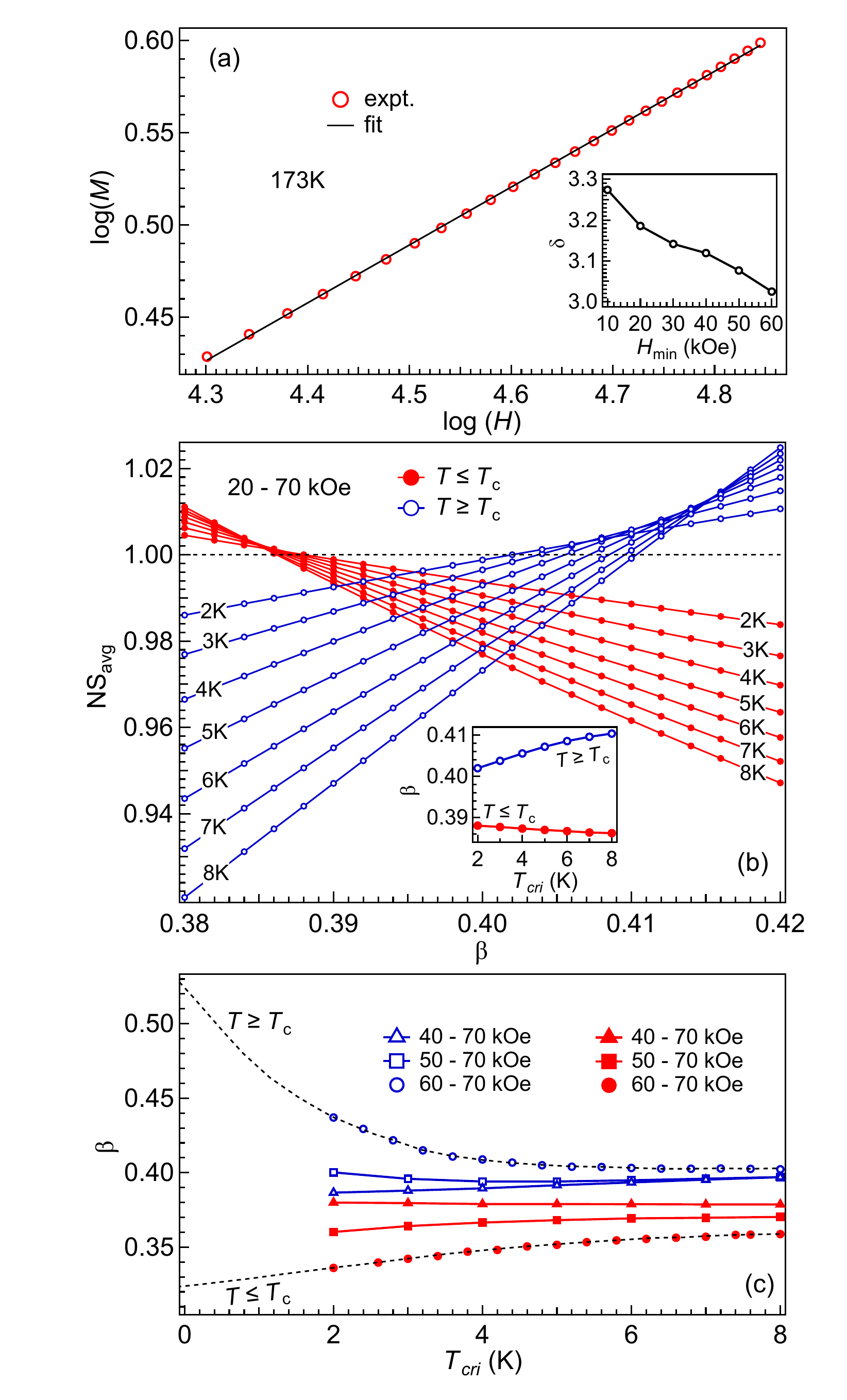}
\caption {(a) The critical isotherm at $T_{\rm c}$ = 173~K on a log-log scale. The black solid line represents the straight fit for $H=20-70$~kOe. The inset shows the dependence of $\delta$ on the minimum value of the magnetic field ($H_{\rm min}$) used to fit the critical isotherm for $H_{\rm max}$ = 70~kOe. (b) The average normalized slope (NS$_{\rm avg}$) as a function of $\beta$ for different $T_{cri}$, both above and below $T_{\rm c}$ = 173~K. The horizontal dashed line represents NS$_{\rm avg}$ = 1. The inset shows the dependence of $\beta$ on the $T_{cri}$ for $T \leqslant T_{\rm c}$ and $T \geqslant T_{\rm c}$. (c) The critical exponent $\beta$ as a function of $T_{cri}$ for different ranges of magnetic fields. The dashed curves are the rough extrapolation of data for $H=60-70$~kOe down to $T_{cri}$ = 0~K, both above and below $T_{\rm c}$. }
\label{Fig3_ANS}
\end{figure}

Recently, an alternative methodology to calculate critical exponents has been proposed, which can be particularly useful for non-trivial systems \cite{Chauhan_arXiv_25}. Employing this approach, the precise value of the transition temperature for Sm$_7$Pd$_3$ is calculated (173.0~K) using the high-resolution low-field dc and ac $M$--$T$ data. Then, the critical exponent $\delta$ is extracted from the slope of the critical isotherm at $T_{\rm c}$ (on a log-log scale) using Eq. (\ref{c3}), as shown in Fig. \ref{Fig3_ANS}(a). Interestingly, we observe a finite dependence of $\delta$ on the range of the magnetic field used to fit this isotherm. The inset of Fig. \ref{Fig3_ANS}(a) shows the field dependence of $\delta$, where the x-axis denotes the lower value of the magnetic field ($H_{\rm min}$) used in the fitting procedure with $H_{\rm max}$ = 70~kOe. The value of $\delta$ decreases with increase in $H_{\rm min}$ and approaches the theoretical value of 3.0 for the mean-field model (see Table I).

Then we manually vary the $\beta$ from 0.38 to 0.42 in steps of 0.002, and calculate the corresponding values of $\gamma$ from the extracted value of $\delta$ for $H=20-70$~kOe ($\delta = 3.186$) using the Widom scaling (WS) relation given as \cite{Widom_JCP_65, Kaul_JMMM_85}

\begin{eqnarray}
 \delta = 1 + \frac{\gamma}{\beta} 
\label{WS}
\end{eqnarray}

For each set of $\beta$ and $\gamma$, the MAP is constructed, and the average normalized slope (NS$_{\rm avg}$) is calculated for $H=20-70$~kOe and plotted as a function of $\beta$, where NS$_{\rm avg}$ is defined as \cite{Chauhan_arXiv_25}
 
\begin{eqnarray}
{\rm NS}_{\rm avg} =\frac{ \sum_{i=1}^{N} ({\rm NS})_i}{N},
\label{NSavg}
\end{eqnarray}

where $i$ represents the NS at different temperatures and $N$ is the number of isotherms under consideration. The NS$_{\rm avg}$ deviates from unity as we move away from the $T_{\rm c}$, and a crossover of NS$_{\rm avg}$($\beta$) = 1 defines the correct value of $\beta$ \cite{Chauhan_arXiv_25}. However, we note that the NS$_{\rm avg}$ calculated for $T \leqslant T_{\rm c}$ and $T \geqslant T_{\rm c}$ gives two different values of $\beta$. It is clear from Eq. (\ref{NSavg}) that the value of NS$_{\rm avg}$ also depends on the temperature regime under  investigation ($T_{cri}$). Therefore, in Fig. \ref{Fig3_ANS}(b), we present the NS$_{\rm avg}$ as a function of $\beta$ for different $T_{cri}$, both above and below $T_{\rm c}$ = 173~K for $H=20-70$~kOe. For $T \geqslant T_{\rm c}$, $\beta_{\rm NS_{\rm avg}=1}$ shows a strong dependence on $T_{cri}$ from 2 (173--175~K) to 8~K (173-181~K), whereas a minimal effect is observed for $T \leqslant T_{\rm c}$ [see Fig. \ref{Fig3_ANS}(b)]. In order to quantitatively understand this, the $T_{cri}$ dependence of $\beta$ is plotted in the inset of Fig. \ref{Fig3_ANS}(b) for $H=20-70$~kOe for both $T \geqslant T_{\rm c}$ and $T \leqslant T_{\rm c}$. This indicates the different rates of the ordering and decay of the magnetization below and above $T_{\rm c}$, and hence different critical regimes on the two sides of the transition in Sm$_7$Pd$_3$. The value of critical exponent $\beta$ and hence the resulting $\gamma$ extracted for $T \leqslant T_{\rm c}$ were proposed as the actual critical exponents in Ref. \cite{Chauhan_arXiv_25}. However, below we demonstrate that both $T \leqslant T_{\rm c}$ and $T \geqslant T_{\rm c}$ are important to estimate the exact values of the critical exponents.  For $T \geqslant T_{\rm c}$, the value of $\beta$ shows a downturn, whereas it shows a small but monotonic upturn for $T \leqslant T_{\rm c}$ as we move deeper into the critical region.

\begin{figure*}
\centering
\includegraphics[width=1\textwidth]{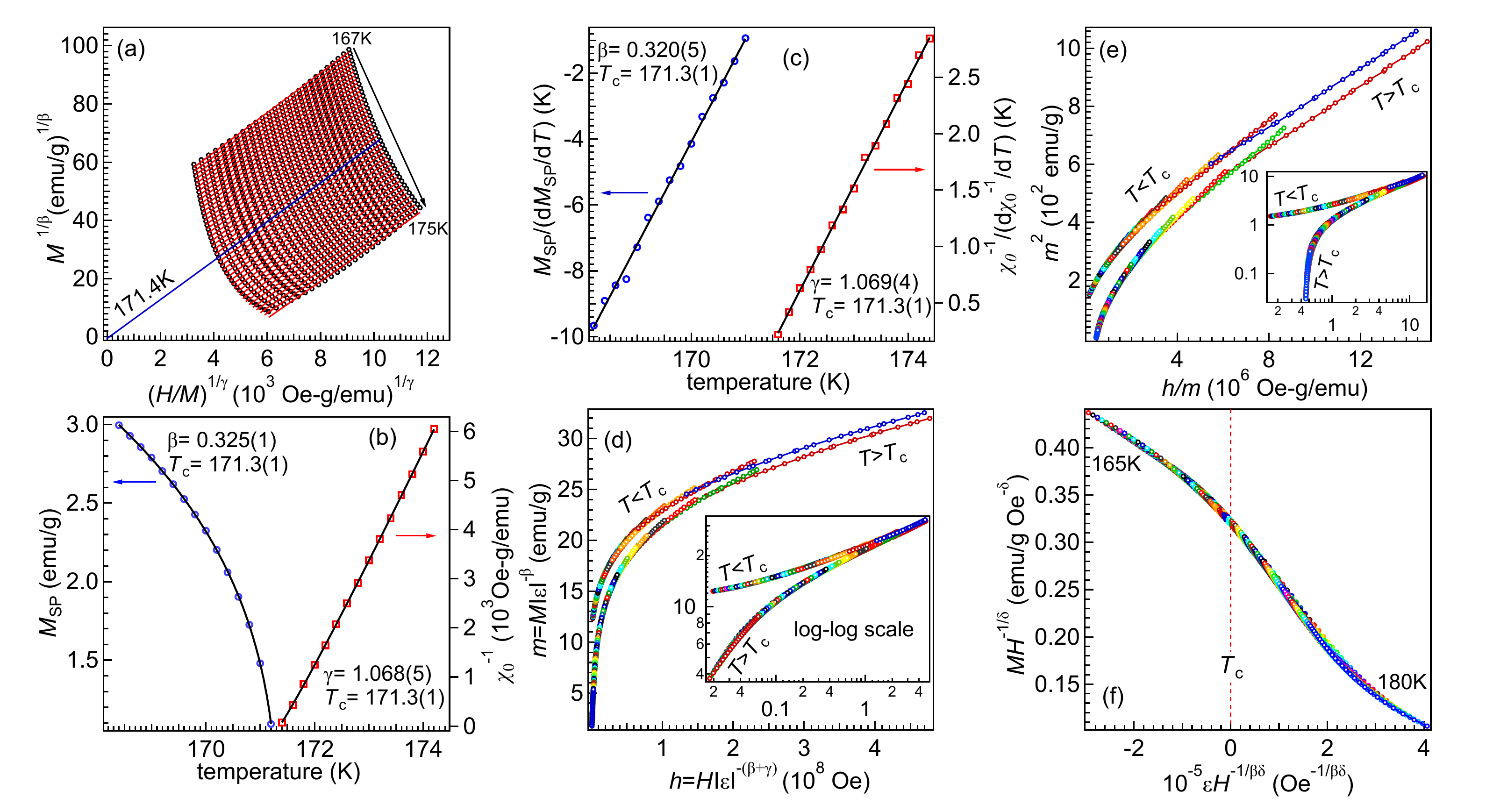}
\caption {(a) Modified Arrott plot constructed using the extrapolated values of the critical exponents. The black circles and red lines show the experimental data points and the linear fit for $H \geqslant $ 20~kOe, respectively, while the blue line represents the extrapolation of the linear fit at $\approx T_{\rm c}$ down to $H=0$. (b) Temperature-dependent spontaneous magnetization $M_{\rm SP}$ (on the left axis) and inverse initial susceptibility $\chi_0^{-1}$ (on the right axis), where the black solid curves represent the best fit using Eqs. (\ref{c1}) and (\ref{c2}), respectively. (c) Temperature-dependent $M_{\rm SP}(T)/[{\rm d}M_{\rm SP}(T)/{\rm d}T]$ (on the left axis) and $\chi_0^{-1}/[({\rm d}\chi_0^{-1})/{\rm d}T]$ (on the right axis) plot, where the black solid lines represent the linear fit using Eqs. (\ref{KF1}) and (\ref{KF2}), respectively. (d) The reduced magnetization, $m$ vs. reduced field $h$, and (e) $m^2$ vs. $h/m$ plot, where the insets show their respective log-log plots. (f) $MH^{-1/\delta}$ vs. $\epsilon H^{-1/(\beta \delta)}$ plot from 165-180~K range, where the vertical dashed line locates the $T_{\rm c}$. } 
\label{Fig4_scaling}
\end{figure*}

We carried out a similar analysis of the critical exponent $\beta$($T_{cri}$) for the different ranges of the magnetic fields, i.e., by varying $H_{\rm min}$ and keeping $H_{\rm max}$ = 70~kOe, using their respective values of $\delta$ [see Fig. S3 of \cite{SI}], as shown in Fig. \ref{Fig3_ANS}(c). Interestingly, unlike the low-field behavior, the value of $\beta$ for $T \leqslant T_{\rm c}$ decreases, whereas that for $T \geqslant T_{\rm c}$ increases as we move closer to the transition, as shown in Fig. \ref{Fig3_ANS}(c). Note that for a given value of $\delta$, Eq. (\ref{WS}) implies that $\beta \propto \gamma$. The increase in the value of $\beta$ (and hence $\gamma$) for $T \geqslant T_{\rm c}$ mimics well with the increase in the value of $\gamma$ with the number of iterations in the standard convergence procedure (SCP), described above [see Fig. \ref{Fig2_SCP}(d)]. Moreover, a decrease in the value of $\beta$ for $T \leqslant T_{\rm c}$ is also consistent with a reduction in the $\beta$ value in SCP [see Fig. \ref{Fig2_SCP}(c)]. This demonstrates that the values of $\beta$ and $\gamma$ must be separately estimated from the crossover of NS$_{\rm avg}$ = 1 line for $T \leqslant T_{\rm c}$ and $T \geqslant T_{\rm c}$, respectively.  It is consistent with the physical interpretation of $\beta$ and $\gamma$, which govern the magnetization of the sample below and above $T_{\rm c}$, respectively. However, this is in contrast with Ref. \cite{Chauhan_arXiv_25}, where both critical exponents $\beta$ and the corresponding $\gamma$ are proposed to be estimated from the $T \leqslant T_{\rm c}$ data only.

The above analysis shows that the values of critical exponents $\beta$ and $\gamma$ are sensitive to both the range of magnetic field ($H_{cri}$) and the temperature ($T_{cri}$) under consideration. Therefore, using Sm$_7$Pd$_3$ as an example, in the following section, we will discuss the importance of properly choosing the $T_{cri}$  and $H_{cri}$ for  the critical behavior analysis of non-trivial magnetic systems. To estimate the correct values of the critical exponents for  Sm$_7$Pd$_3$, we roughly extrapolate the $\beta$($T_{cri}$) down to $T_{cri}$ = 0 for T$\leqslant T_{\rm c}$ as well as T$\geqslant T_{\rm c}$, as shown by the dashed curves in Fig. \ref{Fig3_ANS}(c) for $H=60-70$~kOe, which gives $\beta$ $\approx$ 0.33 and 0.53 (implying $\gamma$ $\approx$ 1.06), respectively. The MAP is constructed using these values of critical exponents, as shown in Fig. \ref{Fig4_scaling}(a). It can be observed that the MAP exhibits almost linear behavior for $H=20-70$~kOe, which suggest the correct (or at least close) estimated values of $\beta$ and $\gamma$, as per the Arrott-Noakes equation. A slight curvature in the curves of the MAP has been observed in the low magnetic field region (not shown), due to the presence of competing magnetic interactions in the sample \cite{Kumar_PRB_25}. The $M_{\rm SP}$ and $\chi_0^{-1}$, extracted from the extrapolation of the high-field region (20$-$70~kOe) of the MAP, are fitted (in $\Delta T$ = $\pm$3~K) using Eqs. (\ref{c1}) and (\ref{c2}), respectively, as shown by the solid black curves in Fig. \ref{Fig4_scaling}(b), which gives $\beta$ = 0.325(1), $\gamma$ = 1.068(5), and $T_{\rm c}$ = 171.3(1)~K.

These values of $\beta$, $\gamma$, and $T_{\rm c}$ have been verified using the Kouvel-Fisher (KF) method \cite{Kouvel_PR_64}, which expresses $M_{\rm SP}$ and $\chi_0^{-1}$ as

\begin{eqnarray}
\label {KF1}
\frac{M_{SP}(T)}{\left(\frac{{\rm d} M_{\rm SP}(T)}{{\rm d}T}\right)} = \frac{T - T_{\rm c}}{\beta} \\
\label {KF2}
\frac{\chi_0^{-1}(T)}{\left(\frac{{\rm d}\chi_0^{-1}(T)}{{\rm d}T}\right)} = \frac{T - T_{\rm c}}{\gamma} \end{eqnarray}

We plot $M_{\rm SP}(T)/[{\rm d}M_{\rm SP}(T)/{\rm d}T]$ (on the left axis) and $\chi_0^{-1}/[{\rm d}\chi_0^{-1}/{\rm d}T]$ (on the right axis) as a function of temperature in Fig. \ref{Fig4_scaling}(c), which exhibits linear behavior as per Eqs. (\ref{KF1}) and (\ref{KF2}). The critical exponents $\beta$ and $\gamma$, extracted from the slopes and $T_{\rm c}$ given by the $y$-intercept of the linear fit [shown by solid black lines in Fig. \ref{Fig4_scaling}(c)], are presented in Table I. These values are in good agreement with those obtained from the MAPs, further indicating the correct extracted values of the critical exponents for Sm$_7$Pd$_3$. Moreover, at $T = T_{\rm c}$, Eqs. (\ref{c1}) and (\ref{c2}) give $M_{\rm SP}$ = $\chi_0^{-1}$ = 0, i.e., the isotherm in MAP at $T_{\rm c}$ (critical isotherm) should pass through the origin. The isotherm at 171.4~K (nearest to $T_{\rm c}$ extracted from the iterative process) is shown by the blue line in Fig. \ref{Fig4_scaling}(a), which passes through the origin, confirming the reliability of the extracted critical exponents.

Although a good agreement between the critical exponents extracted from KF and iterative method is observed, the extracted value of $T_{\rm c}$ from both procedures [171.3(1)~K] is significantly lower than that obtained from the dc and ac susceptibility (173.0~K), which raises questions about the reliability of these exponents. The universal scaling (US) behavior is another indicator  often used to check the accuracy of the critical exponents, which is described by the following scaling equation \cite{Wang_PRB_23}

\begin{eqnarray} 
M(H, \epsilon) = |\epsilon|^\beta f_\pm \left(\frac{H}{|\epsilon|^{\beta+\gamma}}\right), 
\end{eqnarray}

where $f_+$ and $f_-$ denote the regular analytical functions above and below $T_{\rm c}$, respectively. For the correct values of $\beta$ and $\gamma$, the reduced magnetization $m = M |\epsilon|^{-\beta}$ vs. the reduced field $h = \frac{H}{|\epsilon|^{\beta+\gamma}}$ curves should collapse into two distinct master curves for the temperatures above and below $T_{\rm c}$. Figure \ref{Fig4_scaling}(d) presents the $m$ vs. $h$ plot for $H = 20-70$~kOe, which demonstrates the collapse of all curves into two universal branches, as can be more clearly seen from the log-log plot shown in the inset. We use the critical exponents obtained from the KF method to construct these US curves. Additionally, the Arrott plot using reduced magnetization and reduced field ($m^2$ vs. $h/m$) also exhibits two distinct branches above and below $T_{\rm c}$, as shown in Fig. \ref{Fig4_scaling}(e), confirming the validity of the extracted critical exponents $\beta$ and $\gamma$. The inset shows the plot on the log-log scale to magnify the low field ($\sim$20~kOe) regime. Another scaling behavior of the isotherms in the asymptotic region can be expressed as \cite{Liu_PRB_18, Yang_PRB_21}

\begin{eqnarray} 
\frac{H}{M^\delta}= k\left(\frac{\epsilon}{H^{1/\beta}}\right), 
\label{c5}
\end{eqnarray}

where $k(x)$ represents the scaling function. In Fig. \ref{Fig4_scaling}(f), we show the $MH^{-1/\delta}$ vs. $\epsilon H^{-1/(\beta \delta)}$ curves near the $T_{\rm c}$ (represented by the vertical dashed line), which show a nice collapse into a single master curve for $H$ = 20-70~kOe as per Eq. (\ref{c5}), further suggesting the correct choice of the extracted critical exponents for Sm$_7$Pd$_3$.

The extracted values of the critical exponents for Sm$_7$Pd$_3$ using the KF method, validated through the US behavior, are $\beta$ = 0.325(5) and $\gamma$ = 1.068(5). However, these values of the CEs cannot be explained solely by any of the known universal classes (see Table I). The critical exponent $\beta$ indicates the 3D Ising behavior, whereas the value of $\gamma$ is close to the mean field model. Similar non-trivial critical exponents corresponding to two different universality classes below ($\beta$) and above ($\gamma$) the $T_{\rm c}$ have been observed recently in several other compounds and are attributed to the transformation from one universality class to the other across $T_{\rm c}$ (two-sided criticality) \cite{Meng_PRB_23, Chauhan_PRL_22, Gaur_PRB_23, Leonard_PRL_15}. The amorphous Gd$_7$Pd$_3$ \cite{Griffiths_PRB_85} also shows the values of the critical exponents close to Sm$_7$Pd$_3$, which does not alter even with significant hydrogenation of the compound \cite{Griffiths_JPFMP_88} (see Table I for comparison). \par

\begin{figure}
\centering
\includegraphics[width=3.5in]{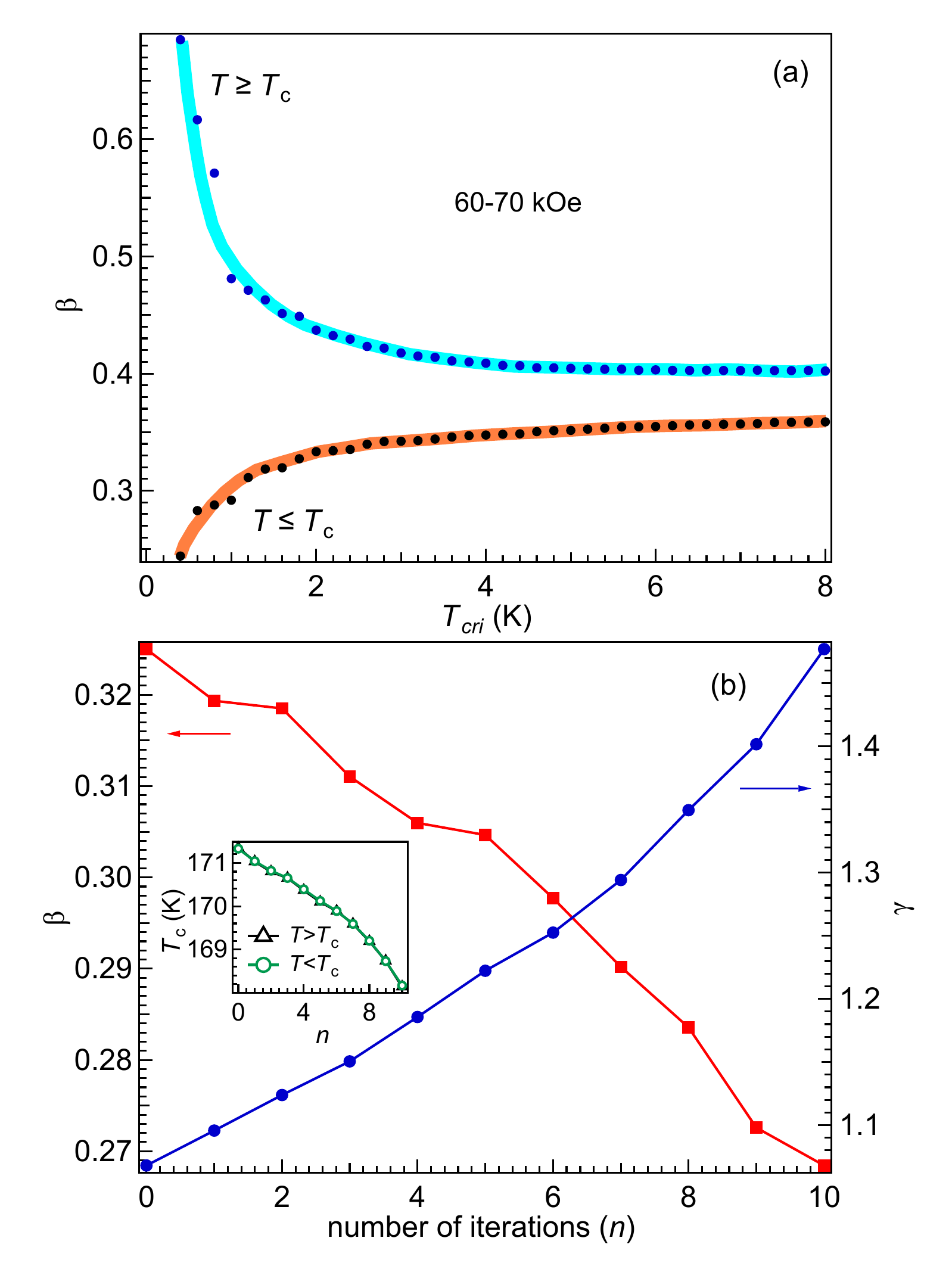}
\caption {(a) The critical exponent $\beta$ as a function of $T_{cri}$ above and below $T_{\rm c}$ for $H = 60-70$~kOe. The shaded curves are the guide to the eyes. (b) The dependence of critical exponents $\beta$ (on the left axis) and $\gamma$ (on the right axis) on the number of iterations. The inset shows the variation in $T_{\rm c}$ with $n$. } 
\label{Fig5_both}
\end{figure}

Here, it is important to reemphasize that the choice of $T_{cri}$ is crucial for a robust critical behavior analysis. Many reports in the literature use temperature steps of $\Delta T \geqslant$ 2~K across the $T_{\rm c}$ for this analysis, and depending on the sharpness of the magnetic transition, the true critical region of the sample can likely be missed. Therefore, in order to more precisely investigate the behavior of Sm$_7$Pd$_3$ in the deep critical region, the value of $\beta$ is calculated using the NS$_{\rm avg}$ method for $H = 60-70$~kOe down to the very close vicinity of $T_{\rm c}$ [see Fig. S3(d) of \cite{SI}]. The critical exponent $\beta$ for the different $T_{cri}$ above and below the $T_{\rm c}$ is plotted in Fig. \ref{Fig5_both}(a) down to  $T_{\rm c}$ $\pm$ 0.4~K. The solid curves are the guide to the eyes. Interestingly, the value of $\beta$ diverges much more rapidly in the critical region than the expected behavior for the $T_{cri}$ $\geqslant T_{\rm c}$ $\pm$ 2~K data discussed above. This behavior clearly suggests that Sm$_7$Pd$_3$ is an unusual case where both critical exponents $\beta$ and $\gamma$ ($\propto$ $\beta$ for T$\geqslant T_{\rm c}$) diverge at $T_{\rm c}$, analogous to the first-order transitions, and no unique value of the critical exponents can explain its magnetism in this region. Note that the diverging nature of the critical exponents is observed in the very close vicinity of the $T_{\rm c}$, predominantly for $T_{\rm c}$ $\pm$2~K, and magnetic isotherms recorded in $T_{cri}$ $\geqslant$ 2~K can result in misleading values of the critical exponents in such cases.

The scaling tests in the critical behavior analysis are the qualitative criteria to determine the accuracy of the critical exponents, and the argument of their collapse or dispersion may depend on the individual choice. To clarify this, we again use the iterative procedure of the MAPs (discussed above), starting from $\beta$ = 0.325 and $\gamma$ = 1.068, the KF results, which give satisfactory scaling behavior. Interestingly, we observe that the critical exponent $\beta$ decreases and $\gamma$ increases monotonically with the number of iterations as shown in Fig. \ref{Fig5_both}(b), consistent with that obtained from the analysis of NS$_{\rm avg}$, discussed above. The critical temperature $T_{\rm c}$ also decreases with the number of iterations, as shown in the inset of Fig. \ref{Fig5_both}(b). This clearly confirms that none of the values of the CEs define the magnetic behavior of Sm$_7$Pd$_3$ in the critical region. Therefore, we propose that along with the well established qualitative scaling tests, one should always present the quantitative change in the critical exponents $\beta$ and $\gamma$ with number of iterations for the more robust critical behavior analysis. 

The presence of non-ferromagnetic (weak antiferromagnetic) components and resulting frustrated magnetic interactions in Sm$_7$Pd$_3$ may contribute to its anomalous behavior in the asymptotic region \cite{Kumar_PRB_25, Silva_PRB_22}. However, this effect is expected to be suppressed in the high-field regime ($H~\geqslant$ 60 kOe) and is therefore less likely to govern the observed non-trivial critical behavior. It is important to recall that Sm$_7$Pd$_3$ exhibits a strong magnetoelastic coupling, where a significant change in lattice parameters accompanies the magnetic phase transition \cite{Biswas_AM_24}. Several other rare-earth compounds with strong magnetoelastic coupling and second-order magnetostructural transitions exhibit similar anomalous critical behavior, which cannot be described by any known universality class \cite{Silva_CM_23, Biswas_Jalcom_22, Halder_PRB_10}.

The influence of spin-lattice coupling on critical behavior near the FM--PM transition has been extensively investigated, both theoretically and experimentally, in systems exhibiting strong magnetoelastic interactions~\cite{Dutta_JPCM_22, Bean_PR_62, Singh_JPCM_19, Callen_PR_63, Mattis_PR_63, PRL_PRL_11, Aharony_PRB_73, Baker_PRL_70}. As early as the 1970s, it was proposed that strong coupling between magnetic exchange energy and lattice degrees of freedom could drive a nominally second-order transition toward first-order behavior~\cite{Bean_PR_62, Callen_PR_63}. Mattis and Schultz, for instance, demonstrated that a magnetothermomechanical first-order transition becomes practically unavoidable when the pressure dependence of the Curie temperature ($\partial T_{\rm c}$/$\partial p$) is sufficiently large \cite{Mattis_PR_63}. In this regard, Sm$_7$Pd$_3$ is an excellent candidate to realize such effects, as it not only displays pronounced magnetoelastic coupling~\cite{Biswas_AM_24} but also shows a strong pressure sensitivity of the Curie temperature, with $\partial T_{\mathrm{c}}/\partial p \approx -9.5$~K/GPa~\cite{Kadomatsu_JMMM_98}. Interestingly, both thermodynamic \cite{Kumar_PRB_20} and magnetization measurements (presented above) indicate a second-order nature of the FM--PM transition in Sm$_7$Pd$_3$. However, the extracted critical exponents exhibit divergent behavior near $T_{\rm c}$, akin to a first-order transition.

Various extensions of the Ginzburg--Landau framework have been proposed to incorporate magnetoelastic coupling, thereby modifying the critical exponents to better capture the unconventional magnetic criticality observed in strongly correlated systems~\cite{Dutta_JPCM_22, Singh_JPCM_19, PRL_PRL_11, Aharony_PRB_73, Baker_PRL_70, Tateiwa_PRB_14}. However, Sm$_7$Pd$_3$ presents an even more complex case, where the critical exponents initially appear to converge to values of $\beta = 0.325(5)$ and $\gamma = 1.068(5)$, as determined from KF analysis and confirmed via scaling behavior. A more detailed examination in the immediate vicinity of $T_{\mathrm{c}}$ reveals, through both the standard convergence procedure and the average normalized slope method, that the critical exponents $\beta$ and $\gamma$ exhibit divergent behavior as $T \rightarrow T_{\mathrm{c}}$: reminiscent of a first-order transition. Therefore, we argue that the strength of spin-lattice coupling, rather than the nominal order of the transition, should dictate the appropriate framework for critical behavior analysis. Our results highlight the need for further theoretical investigations aimed at developing realistic models to explain the critical behavior of strongly correlated magnetic systems with varying degrees of spin-lattice coupling. Until such models are firmly established, caution is advised when applying conventional universality class frameworks to systems exhibiting strong magnetoelastic coupling.


\section{\noindent ~Conclusion}

We have carried out a detailed critical behavior analysis of Sm$_7$Pd$_3$ using the standard convergence procedure (SCP) and the average normalized slope (ANS) method. Both methods reveal that the critical exponents $\beta$ and $\gamma$ diverge near $T_{\rm c}$ = 173 K, indicating non-trivial critical behavior in this system that cannot be described by any established  universality class. This divergence, resembling first-order behavior despite the second-order nature of the transition, originates from the strong magnetoelastic coupling present in the system. These findings not only clarify the magnetic interactions in Sm$_7$Pd$_3$ but also emphasize the broader need for refined theoretical models and analysis techniques in systems where spin-lattice coupling plays a dominant role. By comparing the SCP and ANS methods, we demonstrate that $\beta$ and $\gamma$ should be independently extracted from the $T \leqslant T_{\rm c}$ and $T \geqslant T_{\rm c}$ curves, respectively, when using the latter. Moreover, we show that the accuracy of critical exponents determined via universal scaling behavior serves as a qualitative criterion, and we recommend that the quantitative variation of the exponents with iteration count should be reported when applying SCP. Overall, our results highlight the need for caution when routinely applying conventional critical behavior analysis to systems exhibiting competing magnetic interactions or strong magnetoelastic coupling.

\section*{Acknowledgments}

This work was performed at Ames National Laboratory and was supported by the Division of Materials Science and Engineering of the Office of Basic Energy Sciences, Office of Science of the U.S. Department of Energy (DOE). Ames National Laboratory is operated for the U.S. DOE by Iowa State University of Science and Technology under Contract No. DE-AC02-07CH11358. We thank Mahantesh Khetri for his help in developing code for the critical behavior analysis.

\end{document}